\documentclass[mathleft]{an}
\usepackage{graphicx}
\usepackage{times}
\usepackage{url}
\usepackage{natbib}
\usepackage[modulo,switch]{lineno}
\bibpunct{(}{)}{;}{a}{}{,}
\newcommand{\bibfont}{\small}

\begin{document}

\Pagespan{1}{}
\Yearpublication{2010}
\Yearsubmission{2010}
\Month{1}
\Volume{331}
\Issue{5}
\DOI{10.1002/asna.2010xxxxx}

%
%

\title{Instrument and data analysis challenges\\
    for imaging spectropolarimetry}
\author{C.\ Denker}
\titlerunning{Challenges for imaging spectropolarimetry}
\authorrunning{C.\ Denker}
\institute{
    Astrophysikalisches Institut Potsdam,
    An der Sternwarte 16,
    D-14482 Potsdam,
    Germany}

\received{\today}
\accepted{later}
\publonline{later}

%
%

\abstract{The next generation of solar telescopes will enable us to resolve the
fundamental scales of the solar atmosphere, \textit{i.e.}, the pressure scale
height and the photon mean free path. High-resolution observations of
small-scale structures with sizes down to 50~km require complex post-focus
instruments, which employ adaptive optics (AO) and benefit from advanced image
restoration techniques. The GREGOR Fabry-P\'erot Interferometer (GFPI) will
serve as an example of such an instrument to illustrate the challenges that are
to be expected in instrumentation and data analysis with the next generation of
solar telescopes.}
\maketitle

%
%

\section{Introduction\label{SEC1}}

Ground-based solar observations are currently entering a new era, mainly thanks
to two technological advances: (1) solar AO \citep[see][]{Rimmele2000} to
surmount the deleterious effects of Earth's turbulent atmosphere and (2) an open
telescope design as demonstrated in the Dutch Open Telescope
\citep[DOT,][]{Rutten2004}, which was essential to overcome the aperture
limitation encountered in the traditional design of solar vacuum telescopes. The
New Solar Telescope \citep[NST,][]{Denker2006b} and the German GREGOR project
\citep{Volkmer2007} with apertures of about 1.5~m are currently being
commissioned, thus, paving the way for the 4-meter class Advanced Technology
Solar Telescope \citep[ATST,][entering the construction phase]{Wagner2008} and
the European Solar Telescope \citep[EST,][in the design and development
phase]{Collados2008}.

Imaging spectropolarimeters belong nowadays to the standard equipment of solar
telescopes, since they are photon-efficient and the data can be improved using
image restoration techniques without requiring multi-conjugate AO to observe a
large field-of-view (FOV). This type of instrument has been in use for almost
two decades, \textit{e.g.}, the G\"ottingen Fabry-P\'erot interferometer
\citep{Bendlin1992} and the Telecentric Etalon Solar Spectrometer
\citep[TESOS,][]{Kentischer1998} at the Vacuum Tower Telescope (VTT) on
Tenerife, the Interferometric Bidimensional Spectrometer
\citep[IBIS,][]{Cavallini2006} at the Dunn Solar Telescope (DST) in New Mexico,
and the visible-light and NIR imaging magnetographs \citep{Denker2003a} at Big
Bear Solar Observatory (BBSO) in California. First results of the recently
installed CRISP imaging spectropolarimeter at the Swedish Solar Telescope (SST)
on La Palma were presented in \cite{Scharmer2008a}.

Finally, the transition of the G\"ottingen Fabry-P\'erot interferometer
\citep{Puschmann2006} to the GFPI has been described in \citet{Puschmann2007}.
Early commissioning has already started at GREGOR using a 1-meter aperture CeSiC
mirror on loan from the SolarLite project. After installation of the final
1.5-meter Zerodur mirror commissioning and science demonstration time will
continue until the end of 2011 before entering routine observations in 2012.

Challenges for instrumentation and data analysis will be described in the
following sections focussing on the GFPI. This selection, however, is just a
reflection of the author's bias and familiarity with this instrument. Despite
this, many conclusions should be applicable to other imaging spectropolarimeters
as well as other types of instruments, which are envisioned for the next
generation of solar telescopes.

\begin{figure*}[t]
\centerline{\includegraphics[height=50mm]{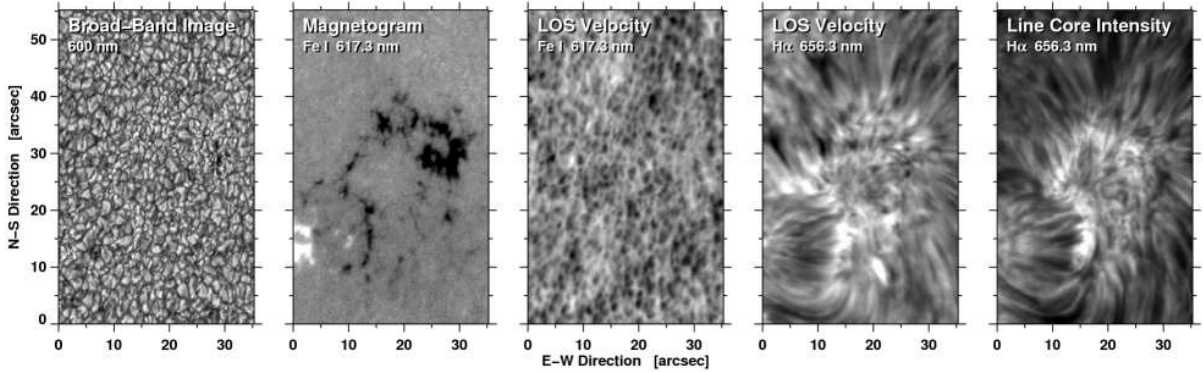}}
\caption{Maps of physical parameters derived with imaging spectropolarimetry for
    a region containing small-scale magnetic fields.}
\label{FIG01}
\end{figure*}

%
%

\section{GREGOR Fabry-P\'erot Interferometer\label{SEC2}}

GFPI is a dual-etalon Fabry-P\'erot interferometer for high-resolution
two-dimensional spectropolarimetry. The coatings of the etalons were optimized
for the wavelength range from 530--860~nm. The spectral resolution is about
${\cal R} = 250,000$, which makes it possible to use spectral line inversion
codes to derive physical parameters from sequences of narrow-band filtergrams.
The image scale is 0.038\arcsec\ pixel$^{-1}$, which results in a FOV of
$52.2\arcsec \times 39.5\arcsec$ taking into account the $1376 \times 1040$
pixel detectors. The diffraction limit of the 1.5-meter GREGOR telescope at
600~nm is $\lambda / D = 0.082\arcsec$, which corresponds to about 60~km on the
solar surface. Image restoration will always be essential to increase the data
quality, since AO correction is only valid for the isoplanatic patch (with a
diameter of $\approx 5\arcsec$) and even with multi-conjugate AO the correction
is limited to FOVs with diameters of $\approx 30\arcsec$. Furthermore,
high-order aberrations might not be completely corrected by AO and MCAO
because of limited bandwidth and wave front reconstruction errors. Image
restoration in combination with MCAO still has to be explored but no principle
problems are to be expected.

Narrow- and broad-band filtergrams are recorded simultaneously at a rate of
about 15~frames~s$^{-1}$ with $2 \times 2$ pixel binning and a 12-bit
digitization depth. Each pair of images is then saved as a binary file using an
on-the-fly compression algorithm. A typical data set consists of $4 \times 8
\times 25 = 800$ images, \textit{i.e.}, four polarization settings, eight images
per wavelength point for speckle polarimetry \citep{Keller1992}, and 25
wavelength points to cover the spectral line. This results in a data volume of
600~MB per line scan. To increase the data rate a RAID-0 controller distributes
the data stream to four SATA harddisks. In this observing mode, the data rate is
about 10~MB~s$^{-1}$. Assuming two hours of observations per day results in
about 75~GB of science data. Approximately the same amount of calibration data
needs to be recorded. The overall data volume of 150~GB per day and a few
terabytes per observing run is still within the limits, which can be handled by
a principle investigator (PI). The data reduction, even though approaching the
limits of office PCs, can still be carried out by the PI, if the results of the
observing program are not immediately needed. An example of physical parameters
derived with imaging spectropolarimetry is shown in Fig.~\ref{FIG01}.

One guiding question for solar instrumentation is, if we make good use of all
available photons. The resolution-lu\-mi\-nosity product of Fabry-P\'erot-based
systems is very good owing to the fact that (1) the transmission of the etalons
is high and that (2) a large FOV is accessible, while only the wavelength domain
needs to be scanned. The resolution-luminosity product is thus an order of
magnitude higher compared to scanning spectrographs employing gratings.
Consequently, only minor gains are to be expected optimizing the optics of the
instrument, AO system and telescope.

One obvious way to improve the photon efficiency of the system is to improve the
duty cycle of the detectors. While the current exposure times are of the order
of 10--20~ms, the data acquisition rate is only 15~frames~s$^{-1}$. However, it
could be as high as 50--100~frames~s$^{-1}$ considering the relatively short
exposure times. Thus, the temporal resolution could be increased 3--6 times
with high frame rate detectors.

Before describing new detector technology in detail, let us briefly discuss the
impact of higher frame rates for image restoration and imaging
spectropolarimetry. Short exposure times ($\Delta t_{\mathrm{exp}} < \tau_0
\approx 40$~ms) are needed for post-facto image restoration to ``freeze'' the
seeing in individual exposures. Here, $\tau_0$ is the typical coherence time of
daytime seeing. Typical exposure times are in the range from 5--20~ms. Assuming
$\Delta t_\mathrm{exp} = 20$~ms, which is a good choice considering that the
narrow-band filtergrams are pho\-ton-starved, we would arrive at a maximum data
acquisition rate of 50~Hz. The other important time scale results from the
photospheric sound speed $c_s \approx 8$~km~s$^{-1}$. The size of a pixel on the
solar surface would be about 28~km, \textit{i.e.}, a feature moving at a
velocity $c_s$ would traverse that distance in about 3.5~s. In principle, only
$3.5 \times 50 = 175$ images could be obtained during this relatively short time
period. If four images are used for full Stokes polarimetry
\citep[see][]{BelloGonzalez2008, Balthasar2009} and 25 wavelength points are
used to cover a line, about two complete scans could be carried out. However,
using speckle polarimetry about eight images are needed per wavelength point,
increasing the observing time four times to about 15~s. In this case, features
traveling faster than 2~km~s$^{-1}$ would violate \textit{Nyquist}'s sampling
theorem. The fact that two images are taken within $\tau_0$ is of minor concern,
since they are obtained in different polarization states and reconstructed
independently.

%
%

\section{Impact of large-format detectors with high data acquisition
    rates\label{SEC3}}

Commercial, of-the-shelf (COTS), large-format detectors\linebreak with high
signal-to-noise (S/N) and high data acquisition rates are currently entering the
market. At the moment the GFPI is using two Imager~QE CCD cameras with Sony
ICX285AL detectors. The cameras are part of a turn-key system manufactured by
LaVison (\url{www.lavision.de}), which runs on the DaVis imaging software. DaVis
also handles the communication with peripheral devices such as the Fabry-P\'erot
etalons, the liquid crystal retarders of the polarimeter, movable mirrors, etc.
COTS technology has the advantage to easily upgrade existing systems once new
(detector) technology becomes available.

Scientific CMOS (sCMOS, \url{www.scmos.com}) is one candidate for future
upgrades of the GFPI, since it will be supported by LaVision's DaVis software.
The format of the sCMOS sensors is $2560 \times 2160$ pixel with a pitch of
6.5~$\mu$m $\times$ 6.5~$\mu$m. A microlens array is used to improve the
photon-collecting efficiency. The wavelength coverage and quantum efficiency
($\mathrm{QE}_\mathrm{max} = 60$\%) is comparable to that of the Imager QE
system. In the global shutter mode using a split frame architecture, the maximum
frame rate is 50~Hz, which is exactly as needed for imaging spectropolarimetry.
Considering that the exposure time is 20~ms and that the charge transfer after
an exposure is complete in less than a 1~$\mu$s, no image smear is expected
using the electronic shutter, \textit{i.e.}, no mechanical shutters are needed.

Anti-blooming ensures that, \textit{e.g.}, bright features such as flares will
not affect neighboring pixels. The rms read noise of the sCMOS detector will be
2--3~electrons for the 50~Hz frame rate. Thus, a dynamic range of 84~dB
(16,000:1) can be reached so that a digitization depth of 14-bit is required to
record the signal without any loss. The intrinsic non-linearity of the sCMOS
device is about 1\% but can be corrected to better than 0.2\%. Dual
analog-to-digital converters (ADCs) with high and low gain settings provide
simultaneously a high S/N and broad dynamic range. This is important for solar
observations considering the low contrast of solar granulation and the high
dynamic range required to observe sunspots. Much of the information contents of
a spectral line is contained close to the core, where the rest intensity is low.
Here, the largest gain of the dual ADCs is to be expected. Finally, the sCMOS
device offers standard features such as binning and region-of-interest (ROI)
read-outs so that read-out speed and/or S/N can be improved.

Carrying out the same computations as in Section~\ref{SEC2}, we arrive at a data
volume of 8.8~GB for the 800 images in a spectral line scan. Note, however, that
even though the data volume is 15 times larger, it only takes a third of the
time to acquire the data. As a benefit, taking the calibration data will only
take a third of time as well, thus reducing the observational overhead by a
factor of three. In summary, making the leap to the next generation of COTS
detector technology will increase the data volume by a factor of 50. In the
following, we will discuss some of the implications and show that this requires
a paradigm shift for ground-based solar instrumentation.

If we assume again a two hours observing day and 30~min worth of calibration
data, the daily data volume will be about 5~TB even with lossless compression.
These data rates and volumes can be handled by todays technology using,
\textit{e.g.}, dual Camera Link interfaces, RAID controllers and SATA-600
harddisks, dual 64-bit/100~MHz PCI buses, and 100~Gigabit Ethernet. These
hardware devices belong to the high-end sector and require knowledgable
technical support to build a working system. Obviously, the resources to deal
with such data volume go well beyond what PIs of an observing run will find at
their disposal in their office.

Similarly, we can arrive at an estimate of the computational efforts required to
analyze such data. Image restoration of a data cube of 800 subimages with $128
\times 128$ pixel takes about 40~s on a single CPU. Speckle deconvolution and
data calibration of narrow-band images raise the computation time for the stack
of 800 subimages to about 80~s. The size of the subimages corresponds
approximately to the diameter of the isoplanatic patch ($\approx 5\arcsec$ or
3600~km on the solar surface). Allowing for some overlap of neighboring
isoplanatic patches, we arrive at $40 \times 34 = 1360$ isoplanatic patches,
which need to be individually restored before being reassembled to yield the
restored spectral line scan. The computation time for one spectral line scan
would be about 30~hours. The total computation time for the 450 spectral line
scans during the two hour observing period would be about 600 days on a single
CPU. Thus, data analysis presents an even more stringent challenge than handling
the data volume. However, using a cluster with 500 quad core CPUs would reduce
the computation time to less than one day as needed to keep up with the stream
of observational data. On a positive note, once the data is reduced, the volume
is reduced by a factor of 10 to about 500~GB per day. While still not small,
this data volume can be handled in the archives of today's virtual
observatories.

%
%

\section{Discussion}

While the previous sections were concerned with the particulars of the GFPI and
possible upgrades to large-format, high frame rate and high S/N detectors, the
discussion will focus on the broader picture. Imaging spectropolarimeters are
photon-efficient instruments and with the advent of new detector technology, the
last remaining inefficiency can be removed. Thus, each precious photon can be
detected. Further gains in photon-efficiency would require multi-in\-stru\-ment,
multi-wavelength observations. This has been implemented in the GREGOR concept
for post-focus instruments. Dichroic beamsplitters (pentaprisms) are used for
simultaneous observations with the GFPI, the scanning near infrared spectrograph
and broad-band imagers.

The Blue Imaging Solar Spectropolarimeter (BLISS) is a 2$^{\mathrm{nd}}$
generation instrument for GREGOR, which will add observing capabilities in the
blue part of the spectrum (below 530~nm) in 20130. This is also cost-effective
considering that the cost of building a telescope scales with the area of the
primary mirror. In terms of spatial resolution, observing at 400~nm instead of
600~nm corresponds to using 2.25-meter instead of a 1.5-meter telescope! On the
downside, multi-instrument observations raise the complexity of instruments,
data calibration/analysis, and operations. Even today data calibration will take
as much (or even more) time than science observations. Therefore, a major effort
has to be spent on the development of a reliable and well documented production
code for data calibration and analysis.

Exponential growth rates are encountered in a variety of areas relevant to
imaging spectropolarimetry: processing speed, harddisk and memory capacity,
network bandwidth (which is today's bottleneck), number of pixel in digital
cameras, but also in the power consumption of computer nodes. The growth rates
are related to what is nowadays called \textit{Moore}'s law, \textit{i.e.}, the
statement that the number of transistors that can be placed at minimum cost on
an integrated circuit has doubled approximately every 18 months
\citep[see][]{Moore1965}. The results of the digital revolution have been of
enormous benefit for astronomical instrumentation.

However, the quest for photon-efficiency has driven the instruments to a level
of complexity, where the data handling, storage and analysis requires high-end
computer technology, \textit{i.e.}, data rates of modern solar instruments reach
the limits of today's technology. Since the next generation of solar telescopes
(ATST and EST) will use even larger detectors, \textit{Moore}'s law only ensures
that the requisite technology will be available. However, we should not expect
that the demand for computer resources will diminish, \textit{i.e.},
photon-efficient instruments will be in step with innovation cycles. Current
predictions project the validity of \textit{Moore}'s law to hold until 2020,
when ATST and EST would be operational.

Annother word of caution might be appropriate at this stage, \textit{Moore}'s
law does not apply to institute budgets or funding schemes! Since data rates and
volume exceed the resources of individual users, the data has to be analyzed in
data centers, which could be located at the host institutions of the post-focus
instruments. This in turn means that significant resources have to be
reallocated or new funding has to be acquired. For new instruments the total
cost of ownership (TCO) has to be considered, which consists of (1) the cost of
acquisition and (2) the cost of operation, which can be full of hidden costs,
\textit{e.g.}, the  personnel and computer resources at the data center. While
COTS parts might help to reduce the cost of an instrument, it has a minor impact
on the cost of operations. The GFPI might be the last of a type of solar
instruments, where TCO could be safely ignored. The operating costs of the next
generation of instruments might be a significant part of the budget of a
national facility.

Imaging spectropolarimeters offer much built-in flexibility, \textit{e.g.},
binning, line selection, polarimetric observing modes, number of wavelength
steps, spectral sampling, \textit{etc.} But does this meet the needs of an
average user? In many cases, the scientific objectives are relatively simple.
Usually, the basic physical parameters of a solar feature need to be measured.
Restricting oneself to some basic settings has the advantage that data base
searches for certain solar phenomena would lead to data sets with similar
characteristics. If data sets become too disjointed, the usability of the
spectropolarimetric data will suffer. Considering the significant investments in
new instruments, a major fraction of observing time should be reserved for
synoptic observations, whereas PI-driven observing campaigns with dedicated
set-ups should be limited to observations, which promise a high scientific
impact. The discussion of how the complex post-focus instruments of the new
generation of solar telescopes can serve a broad community has just begun.

This article is intended as a contribution towards the prioritization of
science, which can be addressed with imaging spectropolarimeters. Efficiency,
reliability, and availability are here the key issues for the respective data
analysis and archiving. The GFPI offers a unique opportunity to explore new
observing paradigms and provides guidance for the next generation of solar
telescopes and instruments.

%
%

\acknowledgements The author expresses his gratitude to Drs. \linebreak
H.~Balthasar, A.\ Hofmann, and J.\ Rendtel for carefully reading the manuscript
and providing ideas, which significantly enhanced the paper. The GFPI is the
latest metamorphosis of Fabry-P\'erot interferometers, which were developed over
the last two decades under the leadership of Dr.\ F.\ Kneer. The 1.5-meter solar
telescope GREGOR is built and operated by the German consortium of the
Kiepenheuer-Institut f\"ur Sonnenphysik in Freiburg, the Astrophysikalisches
Institut Potsdam, and the Max Planck Institut f\"ur Sonnensystemforschung in
Katlenburg-Lindau with contributions by the Institut f\"ur Astrophysik
G\"ottingen and other partners.

%
%


\begin{thebibliography}{17}
\ifx \bisbn   \undefined \def \bisbn  #1{ISBN #1}   \fi
\ifx \binits  \undefined \def \binits#1{#1} \fi
\ifx \bauthor  \undefined \def \bauthor#1{#1} \fi
\ifx \batitle  \undefined \def \batitle#1{#1} \fi
\ifx \bjtitle  \undefined \def \bjtitle#1{#1} \fi
\ifx \bvolume  \undefined \def \bvolume#1{#1} \fi
\ifx \byear  \undefined \def \byear#1{#1} \fi
\ifx \bissue  \undefined \def \bissue#1{#1} \fi
\ifx \bfpage  \undefined \def \bfpage#1{#1} \fi
\ifx \blpage  \undefined \def \blpage #1{#1} \fi
\ifx \burl  \undefined \def \burl#1{#1} \fi
\ifx \binterref  \undefined \def \binterref#1{#1} \fi
\ifx \betal  \undefined \def \betal#1{#1} \fi
\ifx \binstitute  \undefined \def \binstitute#1{#1} \fi
\ifx \bctitle  \undefined \def \bctitle#1{#1} \fi
\ifx \beditor  \undefined \def \beditor#1{#1} \fi
\ifx \bpublisher  \undefined \def \bpublisher#1{#1} \fi
\ifx \bbtitle  \undefined \def \bbtitle#1{#1} \fi
\ifx \bedition  \undefined \def \bedition#1{#1} \fi
\ifx \bseriesno  \undefined \def \bseriesno#1{#1} \fi
\ifx \blocation  \undefined \def \blocation#1{#1} \fi
\ifx \bsertitle  \undefined \def \bsertitle#1{#1} \fi
\ifx \bsnm \undefined \def \bsnm#1{#1} \fi
\ifx \bsuffix \undefined \def \bsuffix#1{#1} \fi
\ifx \bparticle \undefined \def \bparticle#1{#1} \fi
\ifx \barticle \undefined \def \barticle#1{#1} \fi
\ifx \botherref \undefined \def \botherref #1{#1} \fi
\ifx \url \undefined \def \url#1{\textsf{#1}} \fi
\ifx \bchapter \undefined \def \bchapter#1{#1} \fi
\ifx \bbook \undefined \def \bbook#1{#1} \fi
\ifx \bcomment \undefined \def \bcomment#1{#1} \fi
\ifx \oauthor \undefined \def \oauthor#1{#1} \fi
\ifx \citeauthoryear \undefined \def \citeauthoryear#1{#1} \fi
\def \endbibitem {}

\bibitem[\protect\citeauthoryear{{Balthasar} et~al.}{2009}]{Balthasar2009}
\begin{botherref}
\oauthor{\bsnm{{Balthasar}},~\binits{H.}},
\oauthor{\bsnm{{Bello Gonz{\'a}lez}},~\binits{N.}},
\oauthor{\bsnm{{Collados}},~\binits{M.}},
et al.:
2009, in: {Strassmeier}, K.G., {Kosovichev}, A.G., {Beckmann}, J.E. (eds.)
  \textit{Cosmic Magnetic Fields: From Planets, to Stars and Galaxies}, IAU
  Symp. 259, p.~665
\end{botherref}
\endbibitem

\bibitem[\protect\citeauthoryear{{Bello Gonz{\'a}lez} \&
  {Kneer}}{2008}]{BelloGonzalez2008}
\begin{barticle}
\bauthor{\bsnm{{Bello Gonz{\'a}lez}},~\binits{N.}},
  \bauthor{\bsnm{{Kneer}},~\binits{F.}}:
\byear{2008}, \bjtitle{A\&A} \bvolume{480}, 265
\end{barticle}
\endbibitem

\bibitem[\protect\citeauthoryear{{Bendlin}, {Volkmer}, \&
  {Kneer}}{1992}]{Bendlin1992}
\begin{barticle}
\bauthor{\bsnm{{Bendlin}},~\binits{C.}},
  \bauthor{\bsnm{{Volkmer}},~\binits{R.}},
  \bauthor{\bsnm{{Kneer}},~\binits{F.}}:
\byear{1992}, \bjtitle{A\&A} \bvolume{257}, 817
\end{barticle}
\endbibitem

\bibitem[\protect\citeauthoryear{{Cavallini}}{2006}]{Cavallini2006}
\begin{barticle}
\bauthor{\bsnm{{Cavallini}},~\binits{F.}}:
\byear{2006}, \bjtitle{SoPh} \bvolume{236}, 415
\end{barticle}
\endbibitem

\bibitem[\protect\citeauthoryear{{Collados}}{2008}]{Collados2008}
\begin{botherref}
\oauthor{\bsnm{{Collados}},~\binits{M.}}:
2008, in: {Stepp}, L.M., Gilmozzi, R. (eds.) \textit{Ground-Based and Airborne
  Telescopes II}, Proc. SPIE 7012, p.~70120J
\end{botherref}
\endbibitem

\bibitem[\protect\citeauthoryear{{Denker} et~al.}{2003}]{Denker2003a}
\begin{barticle}
\bauthor{\bsnm{{Denker}},~\binits{C.}},
\bauthor{\bsnm{{Didkovsky}},~\binits{L.}},
\bauthor{\bsnm{{Ma}},~\binits{J.}},
et al.:
\byear{2003}, \bjtitle{AN} \bvolume{324}, 332
\end{barticle}
\endbibitem

\bibitem[\protect\citeauthoryear{{Denker} et~al.}{2006}]{Denker2006b}
\begin{botherref}
\oauthor{\bsnm{{Denker}},~\binits{C.}},
\oauthor{\bsnm{{Goode}},~\binits{P.R.}},
\oauthor{\bsnm{{Ren}},~\binits{D.}},
et al.:
2006, in: {Stepp}, L.M. (ed.) \textit{Ground-Based and Airborne Telescopes},
  Proc. SPIE 6267, p.~62670A
\end{botherref}
\endbibitem

\bibitem[\protect\citeauthoryear{{Keller} \& {von der
  L{\"u}he}}{1992}]{Keller1992}
\begin{barticle}
\bauthor{\bsnm{{Keller}},~\binits{C.U.}}, \bauthor{\bsnm{{von der
  L{\"u}he}},~\binits{O.}}:
\byear{1992}, \bjtitle{A\&A} \bvolume{261}, 321
\end{barticle}
\endbibitem

\bibitem[\protect\citeauthoryear{{Kentischer} et~al.}{1998}]{Kentischer1998}
\begin{barticle}
\bauthor{\bsnm{{Kentischer}},~\binits{T.J.}},
  \bauthor{\bsnm{{Schmidt}},~\binits{W.}},
  \bauthor{\bsnm{{Sigwarth}},~\binits{M.}},
  \bauthor{\bsnm{{Uexkuell}},~\binits{M.V.}}:
\byear{1998}, \bjtitle{A\&A} \bvolume{340}, 569
\end{barticle}
\endbibitem

\bibitem[\protect\citeauthoryear{{Moore}}{1965}]{Moore1965}
\begin{botherref}
\oauthor{\bsnm{{Moore}},~\binits{G.E.}}:
1965, Electronics 38
\end{botherref}
\endbibitem

\bibitem[\protect\citeauthoryear{{Puschmann} et~al.}{2007}]{Puschmann2007}
\begin{botherref}
\oauthor{\bsnm{{Puschmann}},~\binits{K.G.}},
  \oauthor{\bsnm{{Kneer}},~\binits{F.}},
  \oauthor{\bsnm{{Nicklas}},~\binits{H.}},
  \oauthor{\bsnm{{Wittmann}},~\binits{A.D.}}:
2007, in: {Kneer}, F., {Puschmann}, K.G., {Wittmann}, A.D. (eds.)
  \textit{Modern Solar Facilities -- Advanced Solar Science}, p.~45
\end{botherref}
\endbibitem

\bibitem[\protect\citeauthoryear{{Puschmann} et~al.}{2006}]{Puschmann2006}
\begin{botherref}
\oauthor{\bsnm{{Puschmann}},~\binits{K.G.}},
  \oauthor{\bsnm{{Kneer}},~\binits{F.}},
  \oauthor{\bsnm{{Seelemann}},~\binits{T.}},
  \oauthor{\bsnm{{Wittmann}},~\binits{A.D.}}:
2006,  \bjtitle{A\&A} \bvolume{451}, 1151
\end{botherref}
\endbibitem

\bibitem[\protect\citeauthoryear{{Rimmele}}{2000}]{Rimmele2000}
\begin{botherref}
\oauthor{\bsnm{{Rimmele}},~\binits{T.R.}}:
2000, in: {Wizinowich}, P.L. (ed.) \textit{Adaptive Optical Systems
  Technology}, Proc. SPIE 4007, p.~218
\end{botherref}
\endbibitem

\bibitem[\protect\citeauthoryear{{Rutten} et~al.}{2004}]{Rutten2004}
\begin{botherref}
\oauthor{\bsnm{{Rutten}},~\binits{R.J.}},
\oauthor{\bsnm{{Bettonvil}},~\binits{F.C.M.}},
\oauthor{\bsnm{{Hammerschlag}},~\binits{R.H.}},
et al.:
2004, in: {Stepanov}, A.V., {Benevolenskaya}, E.E., {Kosovichev}, A.G. (eds.)
  \textit{Multi-Wavelength Investigations of Solar Activity}, IAU Symp. 223,
  p.~597
\end{botherref}
\endbibitem

\bibitem[\protect\citeauthoryear{{Scharmer} et~al.}{2008}]{Scharmer2008a}
\begin{barticle}
\bauthor{\bsnm{{Scharmer}},~\binits{G.B.}},
\bauthor{\bsnm{{Narayan}},~\binits{G.}},
\bauthor{\bsnm{{Hillberg}},~\binits{T.}},
et al.:
\byear{2008}, \bjtitle{ApJL} \bvolume{689}, L69
\end{barticle}
\endbibitem

\bibitem[\protect\citeauthoryear{{Volkmer} et~al.}{2007}]{Volkmer2007}
\begin{botherref}
\oauthor{\bsnm{{Volkmer}},~\binits{R.}},
\oauthor{\bsnm{{von der L{\"u}he}},~\binits{O.}},
\oauthor{\bsnm{{Kneer}},~\binits{F.}},
et al.:
2007, in: {Kneer}, F., {Puschmann}, K.G., {Wittmann}, A.D. (eds.)
  \textit{Modern Solar Facilities -- Advanced Solar Science}, p.~39
\end{botherref}
\endbibitem

\bibitem[\protect\citeauthoryear{{Wagner} et~al.}{2008}]{Wagner2008}
\begin{botherref}
\oauthor{\bsnm{{Wagner}},~\binits{J.}},
\oauthor{\bsnm{{Rimmele}},~\binits{T.R.}},
\oauthor{\bsnm{{Keil}},~\binits{S.}},
et al.:
2008, in: {Stepp}, L.M., {Gilmozzi}, R. (eds.) \textit{Ground-Based and
  Airborne Telescopes II}, Proc. SPIE 7012, p.~70120I
\end{botherref}
\endbibitem

\end{thebibliography}

\end{document}